\documentclass[12pt]{article}
\usepackage{graphicx}
\begin{document}

\centerline {\bf Simple Simulation of Magnetic Structure for Nanoclusters}

\bigskip
Ana Proykova and Dietrich Stauffer$^1$ 

\bigskip
Department of Atomic Physics, University of Sofia,

5 James Bourchier Blvd., Sofia-1126, Bulgaria

\bigskip
$^1$ Institute for Theoretical Physics, Cologne University, 

D-50923 K\"oln, Euroland
\noindent

\bigskip
Abstract:
A simple discrete model 
 for magnetic structures
of chromium nanoclusters, found with the help of
local-spin DFT by Kohl and Bertsch, still confirms
 their conclusion that 
in most of the clusters the magnetic moments are not collinear; instead,
in most of the cases they are oriented within one plane. We also simulate
the destruction of the anti-ferromagnetic ordering by thermal fluctuations.


\bigskip
\bigskip
Nanosized materials posses unique properties, which are intermediate 
between the molecular ones and the properties of the corresponding bulk 
materials.The magnetic behaviour of nanostructured systems is 
generally governed by both the intrinsic properties of the magnetic 
nanoclusters (grains) and the interactions between the grains. The 
present study is related to the clusters themselves, specifically the size 
dependence of the magnetic ordering in chromium and mixed chromium-iron 
clusters. 

This work is based on a simple 
discrete model of the problem known as non-collinear 
magnetic ordering \cite{Car}, and  studied by Kohl and Bertsch 
\cite{kohl} in the frame of a general rotationally invariant form of 
local spin-density theory. 

The revived interest in very small magnetic nanoparticles is mainly caused 
by recent advances in synthesis techniques of nanometer-size magnetic 
particles and their application in magneto-optics 
\cite{Uzdin-PR01,Uzdin,Iva}. For a 
 technological process, it is necessary to 
know the size-evolution of the cluster magnetization in order to 
design materials with a long-lasting property, e.g. high (low) 
magnetic moments, stable structures of particles with specific 
shapes. It worth mentioning that the surface effects of small 
particles introduce two different relaxation times in the system 
\cite{Ani-99} and thus, a dynamical metastability \cite{Ani-Vac}, 
which further increases the interest in having a fast and simple 
tool to compute the property needed.

Bulk chromium crystallizes in  body-centered cubic {\it bcc} structures
\cite{Kittel}
with anti-ferromagnetic ordering of the spins which are collinear
 along a fixed quantization axis. However, in clusters their finite sizes 
lead to a different magnetic behaviour
from the bulk: enhanced moments \cite{Heer,Maria}, an altered temperature
 dependence of the magnetization {\cite{Peter}, and
non-collinear effects {\cite{Car,kohl,Uzdin-CMS,Entel}.

Non-collinear arrangements in anti-ferromagnetic clusters  are now 
obtained with the help of a discrete model of classical spins and Monte Carlo
simulations  \cite{Binder}.

\bigskip
We simplified the computation of Kohl and Bertsch by giving each atom a 
classical Heisenberg spin ${\bf S}$ of fixed magnitude $2 \mu_B$ and restricted
us to an interaction "Hamiltonian" H = $\beta \sum_{i<j}^3 {\bf S_i \cdot S_j}$
with positive (anti-ferromagnetic) exchange coupling $\beta$. A Monte Carlo 
simulation slightly changed every spin's orientation, randomly and independently
of other spins.
The new configuration was accepted if it lowered the configuration energy. 
We started from ten independent random spin configurations to check if several
different energy minima could be found.

\begin{table}
\begin{center}
  \begin{tabular}{||c|c|c|c|c|c|c||}  \hline
   configuration&$-E_H$ &$-E_I$&Diff &$m_H$ &$m_I$ &planar?\\   \hline
    2           &  1    &  1  &  0   &  0   & 0    & linear\\   \hline
    3           &  1    &  1  & 1/2  &  0   & 1/3  & yes   \\   \hline
    4a          &  4    &  4  &  0   &  0   & 0    & linear\\   \hline
    4b          &  3    &  3  &  0   &  0   & 0    & linear\\   \hline
    4c          &  2    &  2  &  0   &  0   & 0    &  no   \\   \hline
    5           &  3.86 &  3  & 0.86 & .114 & 1/5  & yes   \\   \hline
    6           &  6    &  5  &  1   &  0   & 0    & yes   \\   \hline
    12a         & 16.39 & 12  & 4.39 & .057 & 1/6  & yes   \\   \hline
    12b         & 20    & 20  &  0   & .333 & 1/3  & linear\\   \hline
    13          & 18.00 & 12  &  6   & .076 & 1/13 & yes   \\   \hline
    51          &128.00 &128  &  0   & .2549&13/51 & linear\\   \hline
  \end{tabular}

\caption{Cluster properties (H = Heisenberg, I = Ising) for the following
topologies numbered by their number of spins: 2 = single pair,
3 = triangle, 4a = square without 
diagonals, 4b = square with one diagonal, 4c = square with both diagonals, 5 = 
three triangles in one line, 6 = two triangles on top of each other, 12a =
one bcc cube with center site plus 3 adjacent cube centers with all 12 bonds
of the basic cube plus all nearest-neighbour bonds, 12b = 12a with only
nearest-neighbour bonds, 13 = symmetrized 12a with fourth adjacent cube center.
Diff is the energy gain by moving away from collinearity.
51 is a "magic" spherical
 cluster of four shells, containing 8, 6, 12, and 24 chromium atoms
respectively, mimicking  a bcc structure \cite{Ani-98}.
Energies $E$ are measured in units of exchange coupling $\beta$, absolute value
of magnetic moment
per atom, $m$, in units of $2\mu_B$; each atom has a classical spin of 
length $2 \mu_B$. {\bf Diff}
 is the energy difference between Ising and Heisenberg
case. The last column indicates if for the Heisenberg case all
spins point into the $x$-$y$-plane ("planar") or even only along the $x$-axis
("linear").}
\end{center}
\end{table}

\begin{figure}[hbt]
\begin{center}
\includegraphics[angle=-90,scale=0.5]{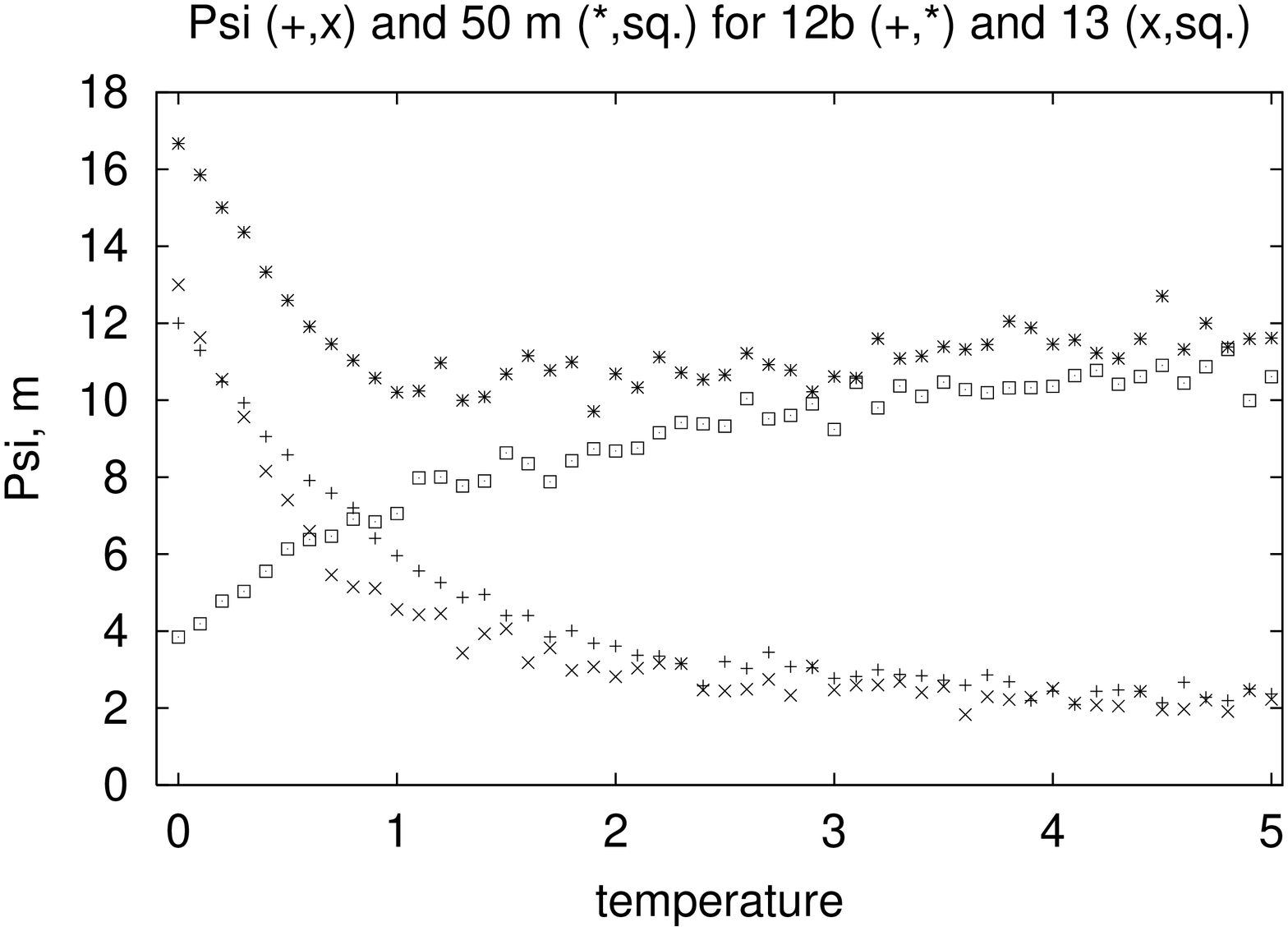}
\end{center}
\caption{Order parameter $\Psi$ (+,x) and absolute value of 
magnetization per atom $m$ (stars, squares;
multiplied with 50 for better visibility), versus $k_BT/\beta$, for 
the completely unfrustrated cluster 12b of 12 spins with nearest-neighbour bonds
(+, stars) and the partially frustrated cluster with 13 spins (x, squares). 
}
\end{figure}

Most of the cluster configurations are shown in Figs. 1 and 3 of Kohl and 
Bertsch and are defined in our table caption. 
Most of the configurations investigated did not use all three directions for 
the spins, and the final spin orientations pointed all in a plane (taken as
the $x$-$y$-plane by fixing (1,0,0) for the first spins and zero for the $z$
component of the second spin). Some of the configurations can be solved exactly
and serve to check the simulation: A single pair has two anti-parallel spins
and an energy $- \beta$. Configuration 3 has its three spins pointing
at angles zero, 120 degrees and 240 degrees from the positive $x$-axis; 4c has
its four spins collinear: right, left, right, left; the two triangles of
6 are oriented opposite to each other, with every triangle as configuration
3; configuration 12b has the top spin right, the four spins in the plane below
all point left, the two in the next lower plane both point right, the four 
in the next lower plane again all point left, and the lowest spin points right.
For 13 spins, each triangle of neighbours has spins as in configuration 
3, as if the other bonds did not exist. For configuration 4c, which could
also be visualized as a tetrahedron, the frustration leads to different
orientations in every simulation, but with the same final energy: Two pairs of
anti-parallel spins, with an arbitrary orientation of one pair to the other,
give the minimum energy of $-2 \beta$ coming from the two unbroken bonds;
the energies of the other bonds add up to zero. Configuration 4b converged very
slowly to its linear ground state when the two spins with only two bonds 
point right, and the two diagonal ones with three bonds point left. 

To simulate mixtures like iron-chromium clusters, we assumed in configuration
12b four of the twelve atoms, forming one horizontal 
plaque of the bcc lattice, to be
connected with their neighbours through ferromagnetic instead of 
anti-ferromagnetic bonds; thus $\beta$ switched sign for them. Now the two 
of the modified spins which have only two neighbours are completely frustrated
and show in various direction for different simulations. The magnetization
is no longer unique, while the energy is always $-11 |\beta|$. 

In four of the above cases collinearity resulted by itself, in the other 
six cases, mostly the larger clusters, it is energetically less favorable. 
The largest cluster of 51 atoms is a "magic" one and because of its special
symmetry it has collinear spins;
omitting a suitable spin destroys both collinearity and planarity.
 'Suitable' is a spin located not on a symmetry axes. 

For collinearity we allow in a separate simulation the spins to show only
to the right or left which defines 
an Ising magnet. The results are also given in the
table. We see that for the larger clusters it may become bad even though
for an infinite bcc lattice it becomes exact. Cheng and Wang \cite{cheng} 
seem to allow zero spins, which we do not allow. (For 51 atoms energy 
minimization did not always find the absolute minimum.)

At any finite temperature $T$, bonds can be broken with a Boltzmann probability
$\propto \exp(-E/k_BT)$ where $E$ is the energy and $k_B$ is the
Boltzmann's constant.
The highest binding energy of the above configuration is obtained for 12b where
all bonds are satisfied. We define an order parameter ("staggered magnetization")
$\Psi$ through the similarity of the spin configuration $S_i$ at finite $T$  
with the ground state configuration $S_i^0$, averaged over many equilibrium
configurations:
$$ \Psi  = \langle | \sum_i {\bf S_i\cdot S_i^0}|\rangle \quad .$$
Standard Metropolis Monte Carlo simulation was applied to this nanocluster 12b.
Fig.1 shows the gradual destruction of magnetic order (Heisenberg model) 
with increasing temperature; of course, no sharp Ne\'el point can be expected 
in such nanoclusters.
Since \cite{cheng}
 the exchange energy $\beta$ is about 2 eV or $\sim 10^4$ Kelvin,
the room temperature corresponds nearly
to the ground state. We averaged over 100 samples of 100,000 Monte Carlo steps
per spin, ignoring the first 90 percent of each sample. We see that the
unfrustrated cluster is more resistant against thermal fluctuations. The
magnetization at high temperatures comes from the random spin orientations
which do not exactly cancel each other.

Our computer simulations, which usually took only seconds, cannot give the
geometrical structure or the magnitude of the magnetic moments; these were 
taken from the much more time consuming calculations of \cite{kohl}. 
No quantitative agreement with  \cite{kohl} could be obtained
because
of our simplification, mentioned already in \cite{kohl}, of only one bond
strength.
Despite this simplification, we confirmed the trend of Kohl and Bertsch that 
clusters of $\sim 10$ atoms
 in general have no collinear spins even though both the smallest
cluster (pair) and the infinite bcc lattice have collinear spins, see the Table. Moreover,
a surprisingly large fraction of the clusters investigated here have all
the spins pointing in the $x$-$y$-plane. Finally, we complemented previous 
works \cite{cheng,kohl} by looking at the thermal destruction of the 
anti-ferromagnetic order in configuration 12b where the ground state satisfies
all bonds.

We thank the Sofia-Cologne university partnership for supporting this 
collaboration, and Hristo Iliev for his help with connectivity
list of the configuration 51.

\end{document}